\documentclass[doublecol]{epl2} 
\usepackage{amsmath}
\usepackage{epsfig}

\graphicspath{{./figures/}}

\title{Reentrant classicality of a damped system}

\shorttitle{Reentrant classicality of a damped system} 

\author{ 
        Benjamin Spreng\inst{1} \and
        Gert-Ludwig Ingold\inst{1}\thanks{E-mail: \email{gert.ingold@physik.uni-augsburg.de}}\and
        Ulrich Weiss\inst{2}}
\shortauthor{B. Spreng \etal}

\institute{         
 \inst{1} Institut f{\"u}r Physik, Universit{\"a}t Augsburg, D-86135 Augsburg,
           Germany
           
  \inst{2} II. Institut f{\"u}r Theoretische Physik, Universit{\"a}t Stuttgart,
           D-70550 Stuttgart, Germany\\
 }

\abstract{For a free particle, the coupling to its environment can be the
relevant mechanism to induce quantum behavior as the temperature is lowered. We
study general linear environments with a spectral density proportional to
$\omega^s$ at low frequencies and consider in particular the specific heat of
the free damped particle. For super-Ohmic baths with $s\ge2$, a reentrant
classical behavior is found. As the temperature is lowered, the specific heat
decreases from the classical value of $k_\mathrm{B}/2$, thereby indicating the
appearence of quantum effects. However, the classical value of the specific heat
is restored as the temperature approaches zero. This surprising behavior is due
to the suppressed density of bath degrees of freedom at low frequencies. For
$s<2$, the specific heat at zero temperature increases linearly with $s$ from
$-k_\mathrm{B}/2$ to $k_\mathrm{B}/2$. An Ohmic bath, $s=1$, is thus very
special in the sense that it represents the only case where the specific heat
vanishes at zero temperature.}

\begin{document}
\maketitle

\section{Introduction}

Physical systems are expected to develop quantum effects as temperature is
lowered. This expectation can be motivated by considering the dimensionless
quantity $k_\mathrm{B}T/\hbar\omega$ where $T$ is the temperature
 and $\omega$ is a characteristic system frequency. Decreasing the temperature
is thus equivalent to an effective increase of Planck's constant $\hbar$ and
consequently to more pronounced quantum effects.  A very particular case in
this respect is the free particle where the system does not possess a frequency
which in our argument could play the role of $\omega$.

Confining the free particle to a finite region of size $L$ supplies
it with a frequency scale $\hbar/2mL^2$. In the following, we  assume that
$L$ is sufficiently large so that finite-size effects are irrelevant at the
temperatures of interest. However, finite-size effects become important as the
strict zero-temperature limit is taken.

Another frequency scale, the one of interest here, arises if we account for the
fact that a physical system is never completely decoupled from its environment.
The coupling to the environment gives rise to damping of the free particle and
provides the damping strength $\gamma$ as a frequency scale. Interestingly, the
dimensionless quantity $k_\mathrm{B}T/\hbar\gamma$ implies that the
environmental coupling drives the particle into the quantum regime, quite in
contrast to the usual expectation that the environment should suppress quantum
effects. The latter would be the case, e.g., for the harmonic oscillator
where the oscillator frequency provides a temperature scale. Our
reasoning might make the free particle appear as very particular.
However, there exist applications where the low-temperature scale is dominated
by the dissipation, as is the case, e.g., for the Casimir effect in the presence
of Drude-type metal boundaries \cite{Ingold09b}.

In order to quantify the properties of the damped free particle, we will focus
on the specific heat. As other thermodynamic quantities, the specific heat of a
system in the presence of a finite coupling to the environment is not uniquely
defined except in the classical limit of very high temperatures
\cite{Haenggi06}. Here, we will define the specific heat on the basis of the
reduced partition function 
\begin{equation}
\label{eq:reducedPartitionFunction}
\mathcal{Z} = \frac{\mathcal{Z}_{\mathrm{S+B}}}{\mathcal{Z}_\mathrm{B} }\, , 
\end{equation}
i.e. the ratio of the partition function of system and bath
$\mathcal{Z}_\mathrm{S+B}$ and the partition function of the bath alone
$\mathcal{Z}_\mathrm{B}$. Based on (\ref{eq:reducedPartitionFunction}) the
specific heat of the damped system is given by the difference
\begin{equation}
\label{eq:specificHeat}
C = C_\mathrm{S+B}-C_\mathrm{B}
\end{equation}
of the specific heats of system and bath on the one hand and of the bath alone
on the other hand. Such a definition is employed, e.g, in the study of impurity
systems \cite{Zitko09,Merker12}.

For an Ohmic environment, it was found that the specific
heat defined in this way can become negative if the coupling between system and
bath is sufficiently strong \cite{Haenggi08}. What may appear as an artifact of
the definition (\ref{eq:specificHeat}) has a clear physical meaning. It can be
shown that the coupling between the free particle and its environment leads to
a shift of environmental modes towards higher frequencies. The resulting
suppression of low-frequency bath modes is responsible for negative values of
the specific heat at low temperatures \cite{Ingold12}.

In the following, we will generalize previous studies of the free particle
coupled to an Ohmic environment to the non-Ohmic case. An example of a
super-Ohmic bath is given by a phononic environment. The specific heat for
non-Ohmic baths has been studied for a harmonic oscillator and related systems
\cite{Ford07,Wang08,Bandyopadhyay10,Wang12}, but as we have seen above, the
damped free particle and the damped harmonic oscillator can behave quite
differently even in their thermodynamic equilibrium properties. In fact,  the
specific heat of the damped free particle at low temperatures has some
surprises to offer.

In the next section we review those aspects of damped systems which will be
required in our study of the specific heat of the damped free particle.  After
providing an expression for the partition function and the specific heat, we
present a qualitative discussion of those features of the dependence of the
specific heat on temperature and damping which are in the focus of the present
paper. Analytical results are then given for the high-temperature quantum
corrections to the specific heat and for its low-temperature behavior. Finally,
we draw our conclusions.

\section{Description of the damped system}

The free damped particle will be modelled by a Hamiltonian in which the system 
described by its position $Q$ and momentum $P$ is
coupled bilinearly to a set of harmonic oscillators with masses $m_n$
and frequencies $\omega_n$ described by their positions $q_n$ and momenta
$p_n$ \cite{Weiss12},
\begin{equation}
\label{eq:hamiltonian}
H = \frac{P^2}{2M} + \sum_{n=1}^{\infty}\left[\frac{p_n^2}{2m_n}
    +\frac{m_n\omega_n^2}{2}\left(q_n-Q\right)^2\right]\,.
\end{equation}
This Hamiltonian contains a potential renormalization term proportional to $Q^2$
in order to ensure that even in the presence of damping a zero mode exists, as is
manifested by the translational invariance of (\ref{eq:hamiltonian}). The parameters
in the Hamiltonian are sufficiently general \cite{Hakim85,Grabert88} to describe
general linear damping because the properties of the damped system depend on the
microscopic parameters of the bath exclusively via the spectral density of bath
oscillators
\begin{equation}
\label{eq:spectralDensityOfBathOscillators}
J(\omega) =
\frac{\pi}{2}\sum_{n=1}^{\infty}m_n\omega_n^3\delta(\omega-\omega_n)\,.
\end{equation}
The total mass $M_\text{bath}$ of the bath oscillators, which will be useful in
the physical interpretation of some of our results, can thus be expressed as
\begin{equation}
\label{eq:m_bath}
M_\text{bath} \equiv \sum_{n=1}^{\infty}m_n = \frac{2}{\pi}\int_0^\infty\text{d}\omega\frac{J(\omega)}{\omega^3}\,.
\end{equation}

The dynamics of the free Brownian particle is specified by the causal velocity
response function $\mathcal{R}(t)$. In the denominator, its Laplace transform
\begin{equation}
\label{eq:velresplapl}
\hat{\mathcal R}(z) = \frac{1}{z+\hat\gamma(z)}
\end{equation}
contains the inertial term as well as the spectral damping function.
The latter is determined by the spectral density $J(\omega)$ through the relation
\begin{equation}
\label{eq:gammaHatSpectralDensity}
\hat\gamma(z) = \frac{2}{\pi M}\int_0^{\infty}\mathrm{d}\omega
                \frac{J(\omega)}{\omega}\frac{z}{\omega^2+z^2}\,.
\end{equation}

To be specific, we assume a continuous distribution of bath oscillators with
a spectral density of the form
\begin{equation}
\label{eq:powerlawSpectralDensity}
J(\omega) = M\gamma\omega^s\frac{\omega_c^{2p-s+1}}{(\omega_c^2+\omega^2)^p}\,.
\end{equation}
For small frequencies, the spectral density (\ref{eq:powerlawSpectralDensity})
increases with the  power law  $\omega^s$. The regimes  $s<1$ and $s>1$
correspond to sub-Ohmic and super-Ohmic damping, respectively, and $s=1$ is the
special case of Ohmic damping. The last factor in
(\ref{eq:powerlawSpectralDensity}) constitutes an algebraic high-frequency
cutoff. The frequency integral (\ref{eq:gammaHatSpectralDensity}) with
(\ref{eq:powerlawSpectralDensity}) is convergent for $s$ in the range
\begin{equation}
\label{eq:range_s_p}
0 < s < 2p+2\,, 
\end{equation}
and yields
\vspace{-2mm}
\begin{equation}
\begin{aligned}
\label{eq:gammaHat}
\hat\gamma(z) &= \frac{\gamma}{\pi}\left(\frac{z}{\omega_c}\right)^{s-1}
               \textstyle{ B(\frac{s}{2}, p+1-\frac{s}{2}) }\\
        &\qquad \times{}_2F_1[\,p,\textstyle{ \frac{s}{2}; p+1; 1-(\frac{z}{\omega_c})^2 }\,]\,. \\
\end{aligned}
\end{equation}
Here, ${}_2F_1$ denotes the hypergeometric function and the beta function can
be expressed in terms of gamma functions as $B(x, y) = \Gamma(x)\Gamma(y)/\Gamma(x+y)$
\cite{NISTHandbook10}.

In the sequel we use units where $\omega_c =k_{\mathrm B} =\hbar=1$. Furthermore, we
will occasionally restrict the range (\ref{eq:range_s_p}) of exponents $s$ to
\begin{equation}
\label{eq:range_s_p_restricted}
0 < s < 2p\,.
\end{equation}
This tighter constraint offers the advantage that in the integral
(\ref{eq:gammaHatSpectralDensity}) the last factor can be replaced by $1/z$ if
$z$ is much bigger than the high-frequency cutoff.

For integer $p$, the $_2F_1$ function can be transformed into a hypergeometric
series which terminates. We then have
\begin{equation}
\label{eq:gamseries}
\begin{aligned}
\hat\gamma(z) =&\frac{\gamma}{(1-z^2)^p}\bigg[\frac{z^{s-1}}{\sin(\frac{\pi s}{2})}\\
  &\!\! + \frac{1}{\pi}\sum_{n=0}^{p-1} \frac{(-1)^{n-1}}{n+1-\frac{s}{2}}
       \frac{B(\frac{s}{2}, p+1-\frac{s}{2})}{B(n+1, p-n)}
       z^{2n+1} \bigg]\,.
\end{aligned}
\end{equation}
This form allows to easily read off the behavior of $\hat\gamma(z)$ for small
and large arguments which will be needed below to determine the specific heat
at low and high temperatures.

The low-temperature properties of the free damped particle, which is the main
focus of this paper, are determined by the low-frequency  characteristics of
$\hat\gamma(z)$. The leading terms in the hypergeometric series in
Eq.~(\ref{eq:gamseries}) yield
\begin{equation}
\label{eq:gam1}
\hat\gamma(z) = \frac{\gamma}{\sin(\pi\frac{s}{2})} z^{s-1}
                +\frac{\Delta M}{M}z
		+\frac{2p+2-s}{4-s}\frac{\Delta M}{M}z^3
		+\mathcal{O}(z^5)
\end{equation}
with
\begin{equation}
\label{eq:massrenorm}
\frac{\Delta M}{M} =  \frac{2\gamma}{\pi}
     \frac{p \, B(\frac{s}{2}, p+1-\frac{s}{2})}{s-2}\,.
\end{equation}
The first term in Eq.~(\ref{eq:gam1}) is responsible for frequency-dependent
damping. The second term adds to the kinetic term $z$ in $\hat{\mathcal R}(z)$.
Its prefactor can therefore be interpreted as an effective change $\Delta M$ of
the particle's mass $M$ due to the coupling to the environment. 
This mass renormalization will be discussed in more detail below.
Finally, the third term only becomes relevant for low temperatures in the
regime $0<s<2$ at the particular point $\Delta M/M=-1$.

In the limit $s\to 2$, both the first  term and the second term in
Eq.~(\ref{eq:gam1}) become  singular. The singularities cancel each other,
however, and a logarithmic term accrues, 
\begin{equation}
\label{eq:gammalog}
\hat\gamma(z) = - \frac{\gamma}{\pi} z[ \, 2 \ln(z) + \psi(p)-\psi(1)\, ] \, .
\end{equation}

\begin{figure}
 \centerline{\includegraphics[width=\columnwidth]{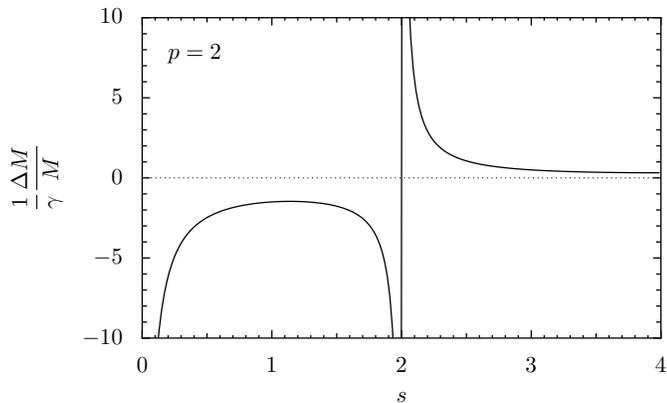}}
 \caption{Mass renormalization (\ref{eq:massrenorm}) for $p=2$ as a function of
 the exponent $s$ of the spectral density of bath oscillators.}
 \label{fig:massrenormalization}
\end{figure}

The relative mass renormalization $\Delta M/M$ is depicted in
Fig.~\ref{fig:massrenormalization} as a function of the exponent $s$ for a
cutoff function characterized by $p=2$.  For $s>2$, the dressed mass $M+\Delta
M$  is larger than the particle's bare mass, since $\Delta M$ is always
positive. From (\ref{eq:m_bath}) and (\ref{eq:gammaHatSpectralDensity}) one
finds that
\begin{equation}
\Delta M = M\hat\gamma'(0)= M_\text{bath}\,,
\end{equation}
where the prime denotes the derivative with respect to the argument. For $s>2$,
we can therefore view the free damped particle as a free particle carrying the
bath oscillators as an extra load around with it. This fact is known from the
long-time behavior of the position autocorrelation function which turns out to
be ballistic \cite{Grabert87}. It will also play a role for the specific heat
at low temperatures.

For $s\leq 2$, the mass (\ref{eq:m_bath}) diverges and the interpretation just
given ceases to hold. However, one can easily convince oneself, e.g. by means of
(\ref{eq:gammaHat}) or (\ref{eq:gamseries}), that for $0<s<2$ the leading term
in (\ref{eq:gam1}) can be obtained as $\hat\gamma(z, p=0)$. The mass renormalization
then becomes
\begin{equation}
\frac{\Delta M}{M} = \lim_{z\to 0}\frac{\text{d}}{\text{d}z}\left(\hat\gamma(z)
-\hat\gamma(z, p=0)\right)
\end{equation}
The interpretation in terms of the mass of bath oscillators again follows from
(\ref{eq:m_bath}) and (\ref{eq:gammaHatSpectralDensity}).  For $0<s<2$, the term
\begin{equation}
\label{eq:deltam_s_lt_2}
\Delta M = M_\text{bath}(p)-M_\text{bath}(p=0) 
\end{equation}
represents the total mass of oscillators which is missing in the actual bath
relative to the reference bath without spectral cutoff. Hence $\Delta M$ is
negative in this range as can also be seen from
Fig.~\ref{fig:massrenormalization}. Indeed, (\ref{eq:deltam_s_lt_2})  describes
the analytical continuation of (\ref{eq:massrenorm}) to the regime $0<s<2$. It
generalizes a result obtained earlier for the special case of Ohmic damping
\cite{Ingold12}.

In the regime $0<s<2$, the contribution $(\Delta M/M) z$ to $\hat\gamma(z)$  is
usually subleading and therefore not of particular interest.  For this reason,
the fact that the renormalized mass $M +\Delta M$  is smaller than the bare mass $M$
largely escaped notice. Most interestingly, the thermodynamic low-energy
properties of the present system are crucially influenced just by the second
term in (\ref{eq:gam1}). More specifically, when $\gamma$ exceeds a
critical value so that $-\Delta M>M$, the renormalized mass $M+\Delta M$ reaches a
negative value which manifests itself as anomalous thermodynamic behavior, as
we shall see.

\section{Reduced partition function and specific heat}

Thermodynamic quantities of the open system are suitably calculated from the
reduced partition function. The partition function of a free particle is only
defined if the particle is confined to a finite region. For a one-dimensional
infinite square well of width $L$ and inner potential $V_0=0$, the energy
levels are given by $E_n = E_{\rm g} \,n^2$ with $E_{\rm g} = \pi^2/2ML^2$, and
the respective partition function ${\mathcal Z}_0 = \sum_{n=1}^\infty\, {\rm
e}^{-(E_\text{g}/T)n^2}$ can be expressed in terms of a Jacobi theta function
$\vartheta_3$ \cite{NISTHandbook10}. In the temperature regime
\begin{equation}
\label{eq:temp0}
 T > E_0 = c E_{\rm g} \, ,
\end{equation}
where $c$ is a positive number sufficiently large so that for $T$ above
$E_0$ the discreteness of the energy eigenstates may be disregarded, the sum
in the partition function can be turned into an integral. We thus arrive
at the classical partition function of the undamped free particle, 
\begin{equation}
\label{eq:zetnull}
\mathcal{Z}_{0,{\rm cl}} = \sqrt{\frac{T}{E_\text{g}}}\int_0^\infty\text{d}x\text{e}^{-x^2}
 =\frac{\sqrt{\pi} }{2}\sqrt{\frac{T}{ E_{\rm g}}}\, .
\end{equation}

For a damped particle, the partition function is augmented by quantum
fluctuations due to the bath coupling. The accessory part may be written as an
infinite Matsubara product \cite{Weiss12} which, under the condition
(\ref{eq:temp0}), does not depend on the width $L$ of the square well. The
resulting reduced partition function reads
\begin{equation}
\label{eq:partfunc}
\mathcal{Z} = \frac{\sqrt{\pi} }{2}\, \sqrt{\frac{T}{ E_{\rm g}}}\,\, \prod_{n=1}^\infty \frac{\nu_n}{\nu_n +\hat\gamma(\nu_n)} \, , \qquad T>E_0 \, ,
\end{equation}
in which $\nu_n =2\pi T n $ are the bosonic Matsubara frequencies.
The subsequent thermodynamic analysis is based on the expression
(\ref{eq:partfunc}).

The specific heat follows from the reduced partition function by
\begin{equation}
\label{specheat1}
C = \frac{\partial}{\partial T} \left( T^2 \frac{\partial \ln(\mathcal{Z})}{\partial T}  \right) .
\end{equation}
Based on the representation (\ref{eq:partfunc}), one then finds
\begin{equation}
\label{eq:specheatmatsu}
C = \frac{1}{2}+\sum_{n=1}^{\infty}\left[
 \frac{[\hat\gamma(\nu_n)-\nu_n\hat\gamma'(\nu_n)]^2}
               {[\nu_n+\hat\gamma(\nu_n)]^2}
   -\frac{\nu_n^2\hat\gamma''(\nu_n)}{\nu_n+\hat\gamma(\nu_n)} 
   \right] \, .
\end{equation}
Here, the prime denotes again the derivative.

A more physical interpretation of (\ref{eq:specheatmatsu}) can be given in terms
of the change $\xi(\omega)$ of the oscillator density caused by the coupling of
the system to the heat bath \cite{Ingold12}. Under the condition
(\ref{eq:range_s_p_restricted}) the Matsubara sum (\ref{eq:specheatmatsu}) can
in fact be rewritten as a frequency integral
\begin{equation}
\label{eq:specheatspectral}
C = \frac{1}{2} + \int_0^\infty {\rm d}\omega\, \xi(\omega)
               \left[C_{\rm ho}(\omega)-1\right]\,.
\end{equation}
Here
\begin{equation}
\label{harmosc}
C_{\rm ho}(\omega) = \left(\frac{\omega}{2T\sinh(\omega/2T)}\right)^2 
\end{equation}
is the  specific heat of a harmonic oscillator with eigenfrequency $\omega$.
The change of the oscillator density can be expressed in terms of
(\ref{eq:velresplapl}) as
\begin{equation}
\label{eq:xiexpr1}
\xi(\omega) = \frac{1}{\pi}{\rm Im}\,\frac{\partial \ln[\hat{\mathcal
R}(-i\,\omega)]}{\partial\omega}\,,
\end{equation}
where $\text{Im}$ denotes the imaginary part.  The fact that this quantity can
take on negative values is at the origin of the peculiar thermodynamic behavior
to be discussed below.

The integral in (\ref{eq:specheatspectral}) vanishes for very high temperatures
so that the specific heat reaches its classical value in this limit. This term
thus represents a pure quantum contribution. On the other hand, at zero
temperature the integral takes the value
\begin{equation}
\label{eq:sumrule}
\int_0^\infty {\rm d}\omega\, \xi(\omega)[-1] = \frac{s-2}{2}\,\Theta(2-s)
\end{equation}
which can also be obtained by taking one half of the zero-frequency limit of the
summand in the Matsubara sum (\ref{eq:specheatmatsu}). As a consequence, the
specific heat tends towards $(s-1)/2$ for $s<2$ while it reaches $1/2$ for
$s\geq 2$.

\begin{figure}
 \centerline{\includegraphics[width=\columnwidth]{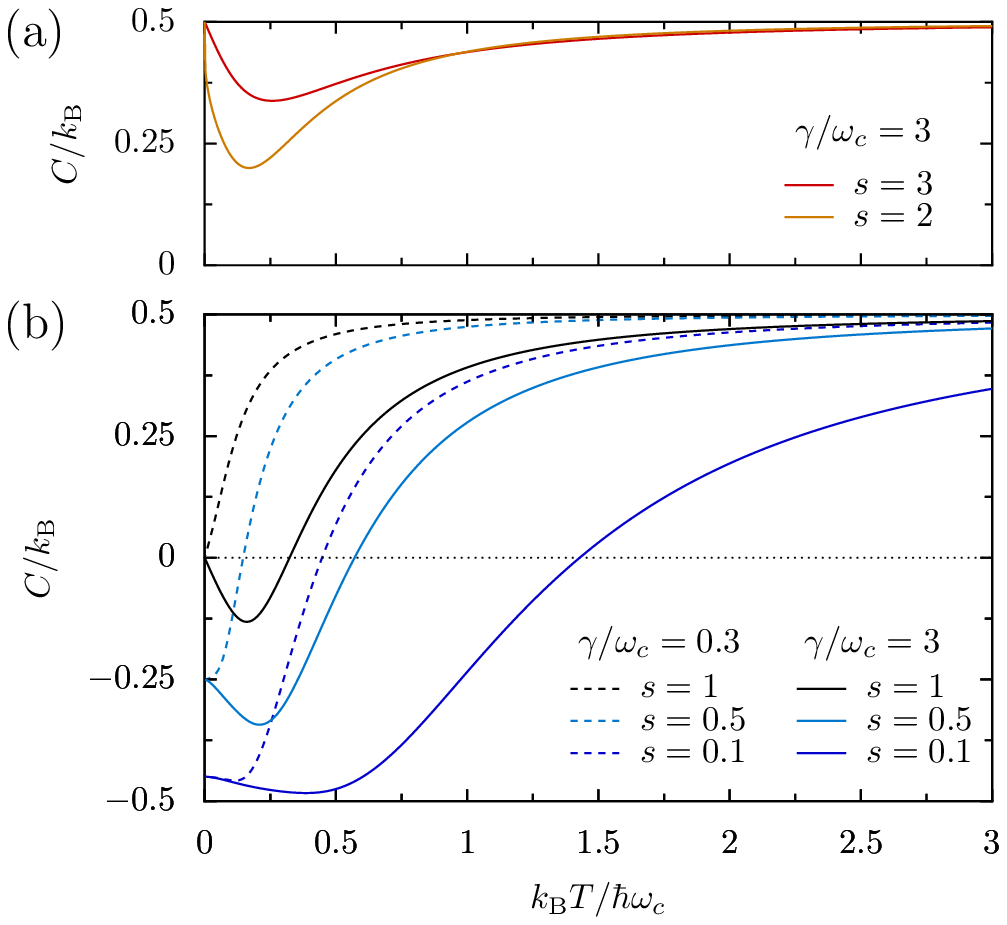}}

 \caption{The specific heat (\ref{eq:specheatmatsu}) of a free damped particle
  is displayed as a function of temperatur for spectral densities of the form
  (\ref{eq:powerlawSpectralDensity}) with a cutoff characterized by $p=2$.
  (a) For $s=2$ and $3$, the phenomenon of reentrant classicality is shown
  for damping strength $\gamma/\omega_c=3$. (b) The appearance of a thermodynamic
  anomaly in the specific heat at low temperatures is shown for $s=0.1, 0.5,$ and $1$
  for $\gamma/\omega_c=0.3$ (dashed curves) and $\gamma/\omega_c=3$ (solid curves).
  For $s=0.5$ and $1$, the dashed curves correspond to a situation where
  $M+\Delta M>0$ so that no thermodynamic anomaly is present
  (cf. Eq.~(\ref{eq:sphlt})). At very low temperatures, the specific heat approaches
  values between $-1/2$ and $1/2$.}
 \label{fig:specificheat}
\end{figure}

Before considering the regimes of low- and high-$T$  where analytical
expressions for the specific heat are available, we discuss the main features of
the temperature dependence of the specific heat obtained from either
(\ref{eq:specheatmatsu}) or (\ref{eq:specheatspectral}).
Figure~\ref{fig:specificheat} displays the specific heat as a function of
temperature for different values of the exponent $s$ with fixed damping
strengths $\gamma/\omega_c=3$ (solid lines) and $0.3$ (dashed lines) and a
cutoff function characterized by $p=2$.

We first discuss the black curve in Fig.~\ref{fig:specificheat}(b)
corresponding to the Ohmic case $s=1$ which depicts a behavior known from the
literature \cite{Haenggi08}. Decreasing the temperature starting from the
high-temperature regime, quantum effects due to the bath coupling lead to a
decrease in the specific heat. It is a particularity of the Ohmic case that at
low temperatures the specific heat tends towards zero. In contrast to the
dashed line the full line corresponds to sufficiently strong damping to allow
for a negative specific heat at low temperatures.  This behavior can be traced
back to a suppression of the density of bath oscillators at low
frequencies due to the coupling to the free particle \cite{Ingold12}. 

Let us now turn to nonohmic damping. With the caveat that zero temperature
actually means $T>E_0$, we see from Fig.~\ref{fig:specificheat}(b) that for
$0<s<2$ the specific heat tends to $(s-1)/2$ in that limit.  For sub-Ohmic
damping, $s<1$, we thus find negative values for the specific heat down to the
lowest temperatures where the partition function (\ref{eq:partfunc}) is still
valid. 

On the other hand, for $s\geq2$ depicted in Fig.~\ref{fig:specificheat}(a), the
specific heat approaches its classical value $k_\text{B}/2$ for very low
temperatures. We thus find the surprising phenomenon of reentrant classicality.
For decreasing temperatures, quantum effects due to the environmental coupling
set in and tend to decrease the specific heat. At very low temperatures, only
the low-frequency oscillators of the environment are relevant but these are
suppressed for $s\geq2$. As a consequence the environmental influence and its
associated quantum effects become less pronounced. Ultimately, the classical
specific heat is restored.  This reentrant behavior is consistent with the fact
mentioned above that for $s>2$ the damped particle behaves in many respects like
an undamped free particle with a renormalized mass. As we had mentioned in the
introduction, a free particle does not possess an intrinsic energy scale and
thus has to behave classically. It should be noted though, that care has to be
taken when measuring the specific heat because the damped free particle
is not ergodic for $s\geq 2$ \cite{Schramm87,Bao05}.

\section{Quantum corrections at high temperatures}

The specific heat at temperatures $T\gg \gamma,\,\omega_c$ is determined by the
behavior of the Laplace transform $\hat\gamma(z)$ of the damping kernel at large
arguments. Provided the exponent $s$ satisfies (\ref{eq:range_s_p_restricted}) 
one finds from (\ref{eq:gammaHatSpectralDensity}) and
(\ref{eq:powerlawSpectralDensity})
\begin{equation}
\label{eq:gamht}
\hat\gamma(z) = \frac{\gamma}{\pi}\frac{B(\frac{s}{2},p-\frac{s}{2})}{z}
         \quad\text{for $z\gg1$}\,.
\end{equation}

The $1/z$ dependence in (\ref{eq:gamht}) implies that for all exponents
$s$ in the range (\ref{eq:range_s_p_restricted}), the leading quantum correction
at high temperatures goes like the square of the inverse temperature
\begin{equation}
\label{eq:sphhtl}
C = \frac{1}{2} - \frac{B(\frac{s}{2}, p-\frac{s}{2})}{12\pi}
  \frac{\gamma}{T^2}
\end{equation}
 
\begin{figure}
\centerline{\includegraphics[width=\columnwidth]{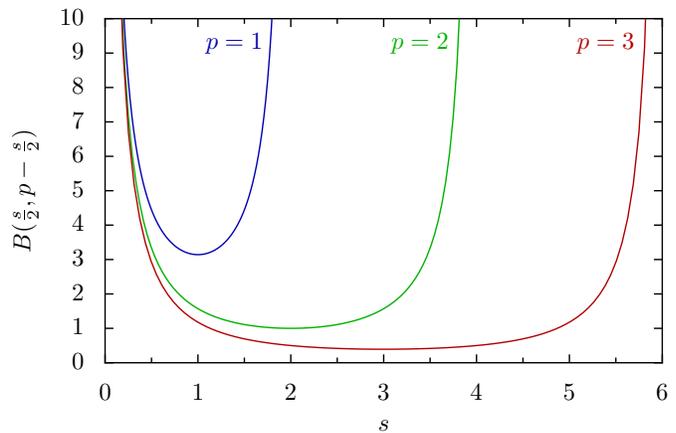}}
\caption{Beta function $B(\frac{s}{2}, p-\frac{s}{2})$ appearing in the large-argument
expansion (\ref{eq:gamht}) of the Laplace transform of the damping kernel as 
a function of the exponent $s$ for cutoffs characterized by $p=1, 2$, and 3.}
\label{fig:betafct}
\end{figure}

The universal $1/T^2$ tail is proportional to the damping strength $\gamma$ and,
after reinserting the constants previously set to one, the cutoff frequency
$\omega_c$. The parameters $s$ and $p$ characterizing the form of the spectral
density of bath oscillators enter via a beta function.  Its U-shaped form
displayed in Fig.~\ref{fig:betafct} possesses a minimum at $s=p$ and
divergencies at the edges $s=0$ and $s=2p$. For a given cutoff function
characterized by $p$, quantum effects can depend sigificantly on the value of
the exponent $s$. For example, in Fig.~\ref{fig:specificheat}(b) quantum effects
at a fixed temperature are significantly larger for $s=0.1$ as compared to the
other values of $s$ for which the specific heat is shown. The quantum
corrections for $p=2$ are weakest for $s=2$.

When, on the other hand, $s$ is kept fixed and the parameter $p$ is increased,
the beta function gets smaller. Thus, sharpening the cutoff function in the
spectral density of bath oscillators
(\ref{eq:spectralDensityOfBathOscillators}) reduces the quantum effects while
the temperature is kept fixed at a large value. 

\section{Low-temperature behavior}

We now study the low temperature regime $E_0  <  T \ll 1, \gamma$.  For $L$
sufficiently large, this domain almost reaches down to zero temperature. The
leading low-temperature correction to the specific heat is obtained from the
leading low-frequency term in the change of the oscillator density
$\xi(\omega)$ defined in (\ref{eq:xiexpr1}). The latter in turn is obtained
from the leading small-argument terms of $\hat\gamma(z)$ according to
(\ref{eq:gam1}) and (\ref{eq:gammalog}).

In the regime $s<2$, the leading contribution to the density
change is found as $\xi(\omega)=a(s, p)\omega^{1-s}$ with 
\begin{equation}  
\label{eq:coefficientA}
a(s, p) = \frac{(2-s)\sin(\pi \frac{s}{2})^2}{\pi\gamma}
              \frac{M+\Delta M}{M}\,.
\end{equation}
This coefficient increases with decreasing damping strength $\gamma$.
Consequently, a reduction of the environmental coupling leads to a more rapid
approach to the classical regime as temperature is increased. We thus recover
the scenario discussed in the introduction according to which damping renders a
free particle more quantum.

From (\ref{eq:specheatspectral}) and the sum rule (\ref{eq:sumrule}), 
we obtain the leading behavior of the specific heat as
\begin{equation}
\label{eq:sphlt}
C= \frac{1}{2}(s-1)+a(s, p)\Gamma(4-s)\zeta(3-s)T^{2-s}\,,
\end{equation}
where $\zeta$ denotes the Riemann zeta function \cite{NISTHandbook10}.
The temperature-independent contribution $\frac{1}{2}(s-1)$ runs from the
negative value  $ -\frac{1}{2}$ reached in the extreme sub-Ohmic limit $s\to
0$ to the classical value $C_{\rm cl}=\frac{1}{2}$ as $s$ is raised to the
super-Ohmic value $s=2$.

Interestingly, only for an Ohmic bath, $s=1$, the specific heat expression
(\ref{eq:sphlt}) is zero at zero temperature, and hence in agreement with the
Third Law, regardless of whether $L$ is finite or infinite. For $s\neq 1$,
the specific heat reaches the nonzero plateau value $\frac{1}{2}(s-1)$ as $T$
approaches the lower bound $E_0$. The drop to zero, $C(T\to 0)\to
0$, and hence accordance with the Third Law takes place only in the
ultra-narrow regime $0\le T<E_0$, in which the discreteness of the energy
spectrum is relevant.

The prefactor of the leading power of $T$ shows a striking anomaly. It changes
sign when the renormalized mass $M+\Delta M$ goes through zero, i.e. when the
damping strength goes through a critical value determined by
(\ref{eq:massrenorm}). Hence the decrease of the specific heat for small
temperatures increasing from zero is a sign that the renormalized mass is
negative. The change from a negative slope at the origin in
Fig.~\ref{fig:specificheat}(b) for the solid lines to a positive slope in
Fig.~\ref{fig:specificheat}(b) for the dashed lines is clearly visible for the
cases $s=0.5$ and $1$.

At the critical point $M+\Delta M=0$, the leading thermal dependence
is ruled by the first and the third term in Eq.~(\ref{eq:gam1}). The
specific heat then reads
\begin{equation}
\begin{aligned}
C&= \frac{1}{2}(s-1)\\
&\quad + \frac{(2+2p-s)\sin(\pi \frac{s}{2})^2}{\pi\gamma}\Gamma(6-s)\zeta(5-s)T^{4-s}\,.
\end{aligned}
\end{equation}
Note that the thermal contribution is consistently positive. 

For the special case $s=2$, in which the expression (\ref{eq:gammalog}) for
$\hat\gamma(z)$ applies,  the fictitious oscillator density takes the form
$\xi(\omega) = - 1/[2\omega \ln^2(\omega)]$, which gives rise to the logarithmic
thermal behavior
\begin{equation} 
C = \frac{1}{2} +\frac{1}{2\ln (T)} \; .
\end{equation} 

Finally, in the regime $s>2$, the second term in the expression (\ref{eq:gam1})
is the leading one, yielding for the density change $\xi(\omega)=-b(s,
p)\omega^{s-3}$ with
\begin{equation}
\label{eq:coefficientB}
b(s, p) = \frac{\gamma}{\pi}(s-2) \frac{M}{M+\Delta M}\,.
\end{equation}
In contrast to the coefficient (\ref{eq:coefficientA}) which was proportional
to $1/\gamma$, the coefficient $b(s, p)$ is proportional to the damping
strength $\gamma$. Nevertheless, since the latter coefficient describes the
quantum corrections to the classical value of the specific heat, the physical
picture is unchanged: Stronger damping renders the free particle more quantum.

The leading contributions to the specific heat for $s>2$ is now obtained as
\begin{equation}
C = \frac{1}{2} - b(s, p)\Gamma(s)\zeta(s-1) T^{s-2}\,.
\end{equation}
Thus, the leading thermal contribution in this regime of exponents $s$ is
always negative.  Again, accordance with the Third Law in the limit $T\to 0$ is
achieved only when the discreteness of the energy spectrum is taken into
account.

\section{Conclusions}

In this Letter, we presented a study of thermodynamic properties of a damped
free quantum particle confined to a box for a general spectral density
(\ref{eq:powerlawSpectralDensity}) of the system-bath coupling. It was found
that the specific heat exhibits surprising features.

In the super-Ohmic regime $s>2$, the specific heat takes the classical value
both at high and at very lemperatures, whereas quantum effects entail reduction
in between. Therefore, reentrant classicality can be observed at low
temperatures.

For $s\neq 1$, i.e. with the exception of the Ohmic case, the third law is
seemingly violated, if we disregard the finite width of the box. The
temperature regime, in which the specific heat eventually goes to zero,
becomes arbitrarily narrow, when the width of the box is chosen correspondingly
large. In the range $0<s<2$ mass renormalization usually plays an ancillary
role. However, in the present model, in which an intrinsic frequency scale is
absent, mass renormalization dictates whether the specific heat at low temperatures
decreases or increases with temperature. The threshold to anomalous behavior
is associated with a vanishing renormalized mass.

\acknowledgments
The authors would like to thank Peter H{\"a}nggi and Peter Talkner
for stimulating discussions. One of us (U.W.) has received financial
support from the Deutsche Forschungsgemeinschaft through SFB/TRR~21.

\end{document}